\documentstyle[twoside,fleqn,espcrc2]{article}

\def\c{\chi}
\def\d{\delta}
\def\g{\gamma}

\def\j{\psi}

\def\l{\lambda}
\def\m{\mu}
\def\n{\nu}
\def\o{\omega}
\def\p{\pi}                     
\def\th{\theta}                  
\def\s{\sigma}                  

\def\G{\Gamma}

\def\O{\Omega}

\def\Th{\Theta}

\def\ca{{\cal A}}

\def\cc{{\cal C}}
\def\cd{{\cal D}}

\def\cf{{\cal F}}

\def\co{{\cal O}}

\def\bo{\raisebox{-.4ex}{\large$\Box$}}                 
\def\cbo{{\,\raise-.15ex\Sc [\,}}                       

\def\sl#1{\rlap{\hbox{$\mskip 1 mu /$}}#1}      
\def\Sl#1{\rlap{\hbox{$\mskip 3 mu /$}}#1}      
\def\braket#1#2{\langle #1 | #2 \rangle}

\def\sbra#1{\left\langle #1\right|}             
\def\sket#1{\left| #1\right\rangle}             
\def\svev#1{\left\langle #1\right\rangle}       

\def\ddt#1{{\buildrel {\hbox{\LARGE .\kern-2pt.}} \over {#1}}}

\def\beqn#1{ \renewcommand{\theequation}{#1}
             \begin{eqnarray} }
\def\eeqn{ \renewcommand{\theequation}{\arabic{equation}}
           \end{eqnarray} }
\def\beqr#1{ \setcounter{equation}{#1}
             \begin{eqnarray} }
\def\bqry{\begin{eqnarray}}
\def\eqry{\end{eqnarray}}
\def\eeqr{\end{eqnarray}}
\def\NON{\nonumber\\}
\def\beqrabc#1{ \setcounter{equation}{0}
                \renewcommand{\theequation}{#1\alph{equation}}
                \begin{eqnarray} }
\def\beqrn#1#2{ \setcounter{equation}{#2}
                \renewcommand{\theequation}{#1.\arabic{equation}}
                \begin{eqnarray} }
\def\beq{\begin{equation}}
\def\eeq{\end{equation}}
\def\seeq#1{eq.~(\ref{#1})}
\def\seEq#1{Eq.~(\ref{#1})}

\def\APH#1{Ann. Phys. {\bf #1}}
\def\CMP#1{Comm. Math. Phys. {\bf #1}}

\def\NPB#1{Nucl. Phys. {\bf B#1}}
\def\NPBP#1{Nucl. Phys. (Proc. Suppl.) {\bf B#1}}
\def\PLB#1{Phys. Lett. {\bf B#1}}
\def\PRD#1{Phys. Rev. {\bf D#1}}

\def\PRL#1{Phys. Rev. Lett. {\bf #1}}
\def\PRP#1{Phys. Rep. {\bf #1}}

\def\APP#1{Acta. Phys. Pol. {\bf B#1}}

\def\MPL#1{Mod. Phys. Lett. {\bf A #1}}

\def\sstyle{\scriptstyle}

\def\rhs{\mbox{r.h.s.} }

\def\ie{\mbox{i.e.} }
\def\eg{\mbox{e.g.} }

\def\frac#1#2{ {\sstyle {#1\over #2} } }

\def\braket#1#2{\langle #1 | #2 \rangle}

\def\tr{{\rm tr}\,}

\def\Re{{\rm Re\,}}

\def\clgt{$\c$LGT}
\def\Heff{H_{\rm eff}(\vec{p}\,)}
\def\fhm{$\j$HM}
\def\Gp{\tilde\G(\vec{p}\,)}
\def\Gpmu{\tilde\G(p_\m)}

\newcommand{\AmS}{{\protect\the\textfont2
  A\kern-.1667em\lower.5ex\hbox{M}\kern-.125emS}}

\title{Lattice Chiral Fermions}

\author{Yigal Shamir\address{School of Physics and Astronomy,
Beverly and Raymond Sackler Faculty of Exact Sciences,\\
Tel-Aviv University, Ramat Aviv 69978, ISRAEL}%
        \thanks{Work supported in part by the US-Israel Binational Science
Foundation, and the Israel Academy of Science.}}

\begin{document}

\begin{abstract}
I review the ongoing attempts to define chiral gauge theories using the
lattice regularization.
\end{abstract}

\maketitle

\section{INTRODUCTION}

  Lattice QCD provides us with a nonperturbative definition of the
Strong Interactions. But so far we do not have a lattice definition for the
complete Standard Model, which is a {\it chiral} gauge theory. Although the
Electro-Weak sector is weakly coupled, there are many important
questions that cannot be resolved by continuum techniques, which are limited
to perturbation  theory or to perturbative expansions around classical
field configurations such as instantons or sphalerons.
To mention a few examples, we need a better understanding of the
Electro-Weak phase transition (see ref.~\cite{jansen} for a recent review)
and of the origin of the net baryon number in
the observed universe. Another deep question has to do with the fact that,
due to triviality, the Higgs sector of the standard model is likely to be only
a low energy effective lagrangian that originates from a more fundamental
theory.

  The goal of constructing chiral lattice gauge theories (\clgt{s} for short)
has not been achieved yet. The basic obstacle is the {\it doubling
problem}~\cite{W,KS,NN}.  In its simplest form, a naive lattice discretization
of the continuum lagrangian of a single Weyl fermion leads to {\it sixteen}
Weyl fermions in the continuum limit. The latter combine into eight Dirac
fermions, thus rendering the continuum limit vector-like.
A general discussion of the conditions for species doubling was first given
by Karsten and Smit~\cite{KS}, who also investigated the relation between the
doublers and the anomaly. A precise mathematical statement of these conditions
was given by Nielsen and Ninomiya~\cite{NN}.

  The Nielsen-Ninomiya theorem asserts that a {\it free lattice
hamiltonian} will have a {\it vector-like} spectrum if certain conditions
hold. In more detail, the assumptions of the theorem include

\begin{itemize}

\item Regular lattice
\item Bilinear hamiltonian
\item Mild form of locality
\item Relativistic low energy spectrum
\item Existence of exactly conserved charges with discrete eigenvalues.

\end{itemize}

\noindent Very briefly, every massless fermion is identified with a two by two
subhamiltonian
\beq
  H_{2\times 2}(\vec{p}\,)=\pm\,\vec\s\cdot(\vec{p} - \vec{p}_c)
  + O((\vec{p} - \vec{p}_c)^2)\,,
\eeq
where $\vec{p}_c$ is the location of the zero in the Brillouin zone, and
$\pm$ defines the chirality. Mild locality is needed to guarantee
{\it continuous first derivatives} for $H(\vec{p}\,)$. The relevant
mathematical theorems then imply the existence of an equal number of
left-handed and right-handed fermions in every complex representation of the
conserved charges.

  From a physicist's point of view, it is not clear why the
No-Go theorems should be insurmountable. In Lattice QCD,
the quarks belong to a complex representation of the non-singlet flavour
symmetries. If we use Wilson fermions, the No-Go theorems are
evaded because all the axial symmetries are broken by the Wilson term.
As a result, exactly conserved axial charges do not exist on the lattice.
However, this non-conservation is just enough to reproduce the
axial anomaly~\cite{KS}, while all the non-singlet
axial symmetries are expected to be restored in the continuum limit.
(I return to this point in Sect.~3.)

  In the case of chiral gauge theories too, one could try to circumvent the
No-Go theorems by invoking a fermion action where gauge invariance is
broken by operators that vanish in the {\it classical} continuum limit.
This approach looks very natural from a perturbative point of view,
and it is the basic idea behind the Smit-Swift~\cite{SS} model.
For example, one can show that the consistent anomaly is correctly
reproduced in lattice perturbation theory~\cite{lptss}.
However, extensive study of the Smit-Swift model at the
nonperturbative level~\cite{noss} has shown that the quantum continuum limit
never leads to a chiral gauge theory. (For a review see ref.~\cite{sseprev}.)
In the symmetric phase(s), the continuum limit is always a {\it vector-like}
theory. This includes realizations where there are no light fermions at all
but only a pure glue theory. In the broken (Higgs) phase
the doublers (typically referred to as ``mirror fermions''~\cite{montvrev}
in this context) can acquire larger masses, but they do not decouple and
remain in the physical spectrum.

  There is a simple and important lesson that should be learned from the
Smit-Swift model. (Other models that exhibit a similar behaviour
will be discussed later.) When the fermion action is not exactly gauge
invariant, the longitudinal component of the
lattice gauge field {\it couples to the fermions}.
This is true even if the perturbative spectrum is anomaly free.
The longitudinal component, which can also be thought of as a
{\it frozen radius Higgs field}, is a strongly fluctuating variable.
The strong fluctuations can, and usually do, change the fermion spectrum,
thus rendering the perturbative analysis highly unreliable.

  This review consists of two parts.
In Sect.~2, I present a unifying framework~\cite{ysnogo}
that clarifies the physical reasons for the
robustness of the No-Go theorems. That framework allows us
to understand why the dynamics of the longitudinal component always gives
rise to a {\it vector-like} spectrum. I begin with the observation that,
even if the fermion action is not gauge invariant, gauge invariance is
restored by the integration over the lattice gauge orbit~\cite{inv,js}.
This kinematical observation, which plays a crucial role, is valid as
long as one uses the standard lattice gauge field's {\it measure}.
Now, invoking asymptotic freedom, the
fermion spectrum of any lattice gauge theory can be determined by setting
$g_0=0$. Switching off the gauge coupling freezes the transversal degrees of
freedom, but not the longitudinal ones. This leads to a fermion-Higgs model
with an {\it exact global symmetry} that corresponds to the original gauge
group.

  The quantum continuum limit of the fermion-Higgs model (or \fhm\ for short)
is a free fermion theory. The spectrum in a given
complex representation can be read off from the zeros
of the inverse two point function $\Gpmu$. At $p_0=0$, $\Gp$
serves as our {\it effective hamiltonian}.
Under extremely mild assumptions on the locality of the action, the effective
hamiltonian $\Gp$ will satisfy all the conditions of the
No-Go theorem, with one important exception. Namely, the No-Go theorem can
be evaded by {\it zeros in the propagator}, or poles in $\Gp$.
One can show~\cite{ghst,pel} that poles in the {\it bilinear part of the
action} are really ghost states that render the theory inconsistent.
However, the analysis of ref.~\cite{ghst,pel} is not directly applicable to
$\Gpmu$, which is an {\it effective action}. It is very important to settle
the question of whether or not this is a real loophole in the No-Go
arguments.

  The second part of this review consists of three sections, each of
which deals with a specific approach to the problem of defining \clgt{s}.
In Sect.~3, I discuss Kaplan's domain wall fermions~\cite{kaplan}, focusing
on two implementations: the waveguide model~\cite{wg}, and overlap
formula of Narayanan and Neuberger~\cite{N&N}. In particular,
I address the question of how to determine the spectrum of the overlap
model nonperturbatively. I also explain how Kaplan fermions might provide
us with a better tool for studying the chiral properties of QCD.

  Sect.~4 deals with the ``gauge fixing'' approach originally introduced
by the Roma group~\cite{roma}. This approach, which was followed
by the Zaragoza group~\cite{zrgz} is so far limited to the context of
lattice perturbation theory. I present various considerations which are
relevant for a nonperturbative implementation of the gauge fixing approach.
The main conclusion is that one should use a {\it global} algorithm,
that selects only relatively {\it smooth} gauge field
configurations~\cite{vink}. I also discuss the relevance of recent work that
reveals a proliferation of Gribov copies on the lattice~\cite{fix}.

  In Sect.~5, I discuss the attempt to give a nonperturbative definition of
chiral gauge theories by putting the gauge fields on the
lattice, while keeping the fermions in the continuum.
Interest in this approach arose following a recent paper by
't Hooft~\cite{thooft}. The bridge between the lattice and the continuum
is provided by a continuum {\it interpolation} of the lattice gauge
field~\cite{intrp}. The lattice is a {\it multiply connected} topological
space. This feature leads to a {\it quasi-local} topological structure
in the interpolating field which, in turn, is the source of certain
difficulties. A short conclusions section ends this review.

\section{THE NO-GO ARGUMENTS}

  In the continuum, whether a given gauge theory is chiral or not depends
on the fermion fields used to define the lagrangian. Thanks to asymptotic
freedom, there has to be a high energy
{\it scaling region} where the lagrangian fields
are the relevant degrees of freedom, and physical processes are well described
by weak coupling perturbation theory. This justifies the identification of
the ``elementary particles'' of the theory with the lagrangian fields.
If we now switch off the gauge coupling we obtain, in the chiral case, a theory
of free fermions with definite handedness in a complex representation of the
gauge group. The latter, in turn, is reduced from a local to a global symmetry.

  On the lattice, there need not be any simple relation between the
lagrangian fields and the elementary fermions of the theory.
(This point will be clarified below.) But at
energies which are well below the lattice cutoff, the above picture
based on continuum physics should hold. We will therefore adopt the same
criterion as in the continuum to determine whether
a given lattice gauge theory is chiral or not.
Namely, we will set $g_0=0$ and ask whether this operation has reduced
the continuum limit to a free theory of chiral fermions.

  Before we proceed, let me mention two caveats. First, the above criterion
ignores the possibility of {\it  composite} gauge bosons. There are some
indications that this could happen in two dimensions~\cite{compW}.
However, the compositeness scenario seems not to work in four
dimensions~\cite{noss}.
Also, while achieving the desired chiral spectrum at $g_0=0$ would certainly
be an important progress, we should be aware that new problems might arise
when the dynamical gauge field is turned back on.

  Setting $g_0=0$ amounts to imposing the constraint
\beq
  \Re\tr(I-U_\Box)=0
\label{g0}
\eeq
on every plaquette. Notice that this constraint {\it does not} imply
$U_{x,\m}=I$ for all links. But \seeq{g0} {\it does} imply that $U_{x,\m}$
is a gauge transform of the identity. Any such {\it pure gauge} configuration
can be written as
\beq
  U_{x,\m}= V_x V^\dagger_{x+\hat\m}
\eeq
Both $U_{x,\m}$ and $V_x$ take values in some compact Lie group $G$.
$V_x$ can be identified with the {\it longitudinal} component of the gauge
field. It can also be viewed a frozen radius Higgs field. Below, I will
often use the notation $I^{(V)}$ to denote pure gauge configurations.

  Thus, setting $g_0=0$ reduces the full gauge field's dynamics to that of
the trivial orbit, which is the collection of all pure gauge configurations.
If the lattice action is gauge invariant, the $V_x$ field decouples and we
are allowed to set $U_{x,\m}=I$ without loosing anything. But proposals for
\clgt{s} often invoke a fermion action which is {\it not} gauge invariant.
In this case,  setting $g_0=0$ leaves behind a \fhm\ (fermion-Higgs
model) where the $V_x$ field interacts with the fermions.

  A crucial feature is that the group $G$ always turns out to be an
{\it exact global symmetry} of the \fhm.
The reader may wonder how this could happen, since we started with
a fermion action which is not gauge invariant. The answer is that because of
the local gauge invariance of the {\it lattice gauge field's measure},
we can rewrite a lattice gauge theory with an arbitrary action
as a lattice gauge theory with a {\it gauge invariant} action, at the price
of introducing the $V_x$ field explicitly. Admittedly, $V_x$ is a
{\it constrained} scalar field, but constrained
scalar fields are perfectly legitimate on the lattice.

  In order to see the emergence of the global symmetry $G$
let us arrive at the \fhm\ in a somewhat
different way~\cite{inv,js}. Consider a lattice gauge theory defined by the
partition function
\beq
  Z = \int \cd U  \cd \j  \cd \bar\j\, e^{-S(U,\j,\bar\j)} \,,
\label{z}
\eeq
\beq
S(U,\j,\bar\j) = S_G(U) + S_F(U,\j,\bar\j)\,.
\eeq
In writing the measure I have used the shorthand
$\cd U = \prod_{x,\m} dU_{x,\m}$ etc.
For simplicity I will assume that $S_G(U)$ is the standard plaquette action.
I make no assumptions on the fermion action $S_F$ which
may or may not be gauge invariant. Apart from gauge interactions, $S_F$
may contain additional multi-fermion or Yukawa interactions.
(However, additional scalar fields are suppressed since their
presence changes nothing in the following arguments.)

  Consider now the effect of a gauge transformation parametrized by $V_x$.
The lattice gauge field transforms according to
\beq
  U_{x,\m} \to U_{x,\m}^{(V)} = V_x U_{x,\m} V^\dagger_{x+\hat\m} \,.
\eeq
A prototype {\it gauge variant} fermion action is
\beq
  S_F = S_K(U) + S_W\,,
\label{sfss}
\eeq
\beq
  S_K = \sum \bar\j (\sl\partial P_R + \Sl{D}(U) P_L) \j \,,
\label{sk}
\eeq
and $S_W$ is the (free) Wilson term.
The gauge variance of $S_F$ means that there is no obvious choice for
the gauge transformation of the fermion variables.
For example, we may decide to apply the transformation only to the
left-handed part of the fermion field, \ie
\bqry
  \j_{Lx} & \to & \j^{(V)}_{Lx} = V_x\j_{Lx} \,, \NON
  \j_{Rx} & \to & \j^{(V)}_{Rx} = \j_{Rx} \,,
\label{trans}
\eqry
and similarly for $\bar\j_{Lx}$ and $\bar\j_{Rx}$.
This choice leaves $S_K$ (but not $S_W$) invariant.

  A change of variables in the partition function now leads to
\bqry
  Z & = & \int \cd U  \cd \j  \cd \bar\j\,
         e^{-S(U^{(V)},\j^{(V)},\bar\j^{(V)})}
\label{zz} \\
   & = & \int \cd U \cd V  \cd \j  \cd \bar\j \,
         e^{-S(U^{(V)},\j^{(V)},\bar\j^{(V)})}
\label{zzz}
\eqry
The equality of \seeq{z} and \seeq{zz} follows from
the invariance of the lattice measure under gauge transformations,
and in going from \seeq{zz} to \seeq{zzz} one averages over
all gauge transformations ($\cd V = \prod_x dV_x$).

  We now observe that the new action, given explicitly by
\beq
  S_G(U) + S_F(U^{(V)},\j^{(V)},\bar\j^{(V)}) \,,
\eeq
is gauge invariant! Under a gauge transformation parametrized by $g_x$, the
original gauge and fermion fields transforms as before with $g_x$ replacing
$V_x$, \eg $U_{x,\m} \to U_{x,\m}^{(g)}$,
whereas the new field $V_x$ transforms according to
$V_x \to V_x^{(g)} = V_x g^\dagger_x$.

  The above trick is so simple that the result might look suspicious.
One may worry that the gauge invariance of the new action is ``fictitious'',
if originally $S_F$ was not gauge invariant. I believe that this is not the
case. The transition from \seeq{z} to \seeq{zzz} shows that in
a lattice gauge theory,
{\it a local separation of the longitudinal and transversal degrees of
freedom is always possible, and the transversal degrees of freedom
always couple to a conserved current}.

  Since the new action is gauge invariant, switching off the gauge coupling
is the same as setting $U_{x,\m}=I$ for all links in \seeq{zzz}.
We thus arrive at the \fhm\ defined by the partition function
\beq
  Z' = \int \cd V  \cd \j  \cd \bar\j \,
         e^{-S'(V,\j,\bar\j)}\,,
\label{z'}
\eeq
\beq
  S'(V,\j,\bar\j)=S_F(I^{(V)},\j^{(V)},\bar\j^{(V)}) \,.
\eeq
Notice that $S_G$ drops out because it is $V$-independent.
For $S_F$ of \seeq{sfss}, the resulting \fhm\ is the Smit-Swift model.

  What happens if we choose a different transformation law for the fermion
variables? For example, instead of \seeq{trans}, we may decide to leave
{\it all} the fermion variables inert. This would lead to a different
action $S''=S_F(I^{(V)},\j,\bar\j)$ for the \fhm. However,
$S''$ reduces to $S'$ if we make the field redefinition $\j_L\to \j_L'=V\j_L$
(and similarly for $\bar\j_L$). Since  the field redefinition is
unitary, the two partition functions with actions $S'$ and $S''$
define {\it the same} \fhm.

  As promised, the gauge invariance of $Z$ when expressed in terms of the
additional $V_x$ field, translates into an {\it exact global symmetry} $G$
of $Z'$. If $G$ is spontaneously broken, we arrive at a {\it mirror
fermion} model~\cite{montvrev}. In this review I will be interested
in genuine \clgt{s}  only, so I will restrict my attention to
{\it symmetric} phases.

  Typically, there will be one or two symmetric phases. Let us focus on one
of them, and ask whether the fermion spectrum is chiral or
vector-like with respect to the global symmetry $G$. Considerable
simplification occurs because the quantum continuum
limit is a theory of {\it free massless fermions}.
Here I am assuming the absence of exactly massless scalars.
While a chiral (massless) fermion spectrum should be stable against
infinitesimal modifications of the action, the existence of a
massless scalar in a symmetric phase of the \fhm\ is always {\it accidental}.
Allowing for small modifications of the action, I can therefore assume
the absence of exactly massless scalars without loss of generality.

  Consider now a specific complex representation $\cc=\cc(G)$.
Let $\c_i=\c_i(V,\j,\bar\j)$ be a set of local fermion operators that belong
to $\cc(G)$, which create all the massless fermions in $\cc(G)$ (if there
are any). The $\c_i$-s should be chosen such that
every $\c_i$ creates at least one massless fermionic state when acting on
the vacuum. Such an economic set will always contain a finite number of
$\c_i$-s. Because of the freedom in making field
redefinitions that involve the $V_x$ field,
there need not be one-to-one correspondence between the
lagrangian fields and the asymptotic states of the \fhm.
By taking tensor products of $V$ times an odd number of fermion fields,
we can build operators that belong to practically every representation of $G$.
The question of which of these operators create massless fermions
is clearly a {\it dynamical} one.

  Given the $\c$-s (from now on I suppress all indices)
we calculate the inverse two point function defined by
\beq
  \G^{-1} = \svev{\c(x)\,\c^\dagger(y)} \,.
\eeq
Let us now consider the Fourier transform $\tilde\G(p_\m)$
at $p_0=0$. The crucial observation is that $\tilde\G(\vec{p}\,)$ can serve as
an effective hamiltonian, to which the considerations of the No-Go theorem
can be applied~\cite{ysnogo}.
A sufficient condition for hermiticity of $\tilde\G(\vec{p}\,)$ is
the existence of a spectral representation. Even more generally one can
argue that, in a consistent theory, the massless fermion spectrum must
be determined by the zeros of the hermitian part of $\tilde\G(\vec{p}\,)$,
denoted $\Heff$ from now on. If $\tilde\G(\vec{p}\,)$ had an
anti-hermitian part that does not vanish at  a zero of
$\Heff$, this would mean that a zero energy fermion has a finite
probability to decay, which is clearly a pathological situation.

  At this stage we have a hermitian matrix $\Heff$ which is
a function of the lattice momentum, whose zeros are in one-to-one
correspondence with the massless fermions that belong to $\cc(G)$.
The No-Go theorem will apply, and the spectrum will be vector-like,
if $\Heff$ has continuous first derivatives.

  In order to determine the analytic
structure of $\Heff$ we have to consider three physically
distinct regions~\cite{ysnogo}. The first region is the vicinity of
{\it zeros} of $\Heff$. The second (possible) region is the vicinity
of {\it poles} of $\Heff$, or zeros of the
propagator. The third region covers the rest of the Brillouin zone.
For the moment let me assume that there are no poles in $\Heff$.
Under extremely mild assumptions on the locality of the action,
one expects that $\Heff$ will have continuous first
derivatives away from its zeros. If the action contains only
short range couplings, $\Heff$ should be an {\it analytic}
function of $\vec{p}$ away from the zeros. Thus, what remains is to
establish the continuity of the first derivatives {\it at} the zeros.

  As explained earlier, one can assume the absence of exactly massless scalars.
As a result, the quantum continuum limit in a symmetric phase of the \fhm\
is a theory of free massless fermions. At energies which are
small compared to the mass of the {\it lightest massive excitation},
the effective low energy lagrangian can
contain only non-renormalizable interactions,
the first possible non-linear term being a four-fermion interaction.
Now, a logarithmic term in $\Heff$ can arise only from
diagrams that contain at least two vertices.
Non-renormalizable interactions always have dimensionful coupling constants,
and when these coupling constants occur in front of a logarithmic term
in $\Heff$, they will be multiplied by at least two powers of $p^2$.
This ensures the continuity of the first derivatives
of $\Heff$ at the zeros.

  It remains to consider the possibility of poles in $\Heff$.
Poles in $\Heff$ are, tentatively, the most important
loophole in the No-Go arguments. Their existence cannot be
easily ruled out by locality arguments, because these arguments
usually apply to the {\it propagator}, and not to the inverse
propagator. A pole in the inverse propagator means a zero in the propagator,
and the latter is compatible with analyticity.

  The danger is that poles in $\Heff$
actually describe {\it massless ghost states} (bosonic spinors)
that couple to the gauge field, leading to an inconsistent
theory~\cite{ghst,pel}. At an intuitive level, this can be understood
by considering a bilinear fermion action with a pole at $\vec{p}=\vec{p}_c$.
The contribution to the fermion determinant from momentum eigenstates in the
vicinity of $\vec{p}_c$ can be reproduced by a bosonic field with a first
order action.

  A non-local two component lattice action, in which the fifteen
extra zeros (doublers) were traded with poles,
was proposed some ten years ago by Rebbi~\cite{rebbi}.
A calculation of the vacuum polarization reveals the inconsistency of
this proposal~\cite{ghst}. The coefficient of the logarithmic term
turns out to be $(-14)$ times the right answer. The explanation is that
$-14=1-15$, \ie the fifteen poles {\it do} contribute to the logarithmic term,
and with the {\it wrong} sign.
Moreover, Pelissetto~\cite{pel} showed that this phenomenon is completely
general, and it occurs whenever the extra zeros are traded with poles,
regardless of the detailed form of the action.

  The crucial ingredient of the analysis of ref.~\cite{ghst,pel}
is the Ward identity relating the gauge field's vertex to the inverse
propagator. In addition, the analysis relies on
standard diagrammatic rules for calculating the vacuum polarization
in terms of the vertex and the propagator. The calculation proceeds by
showing that the poles in the inverse propagator reappear in the vertex,
and make an important contribution in the limit of small external
momentum.

  The difference between ref.~\cite{ghst,pel} and our general setup
is that the inverse propagator $\tilde\G(p_\m)$ is an {\it effective action}.
Consequently, a simple diagrammatic relation between $\tilde\G(p_\m)$,
the vertex function and the vacuum polarization is lacking.
However, as the following argument suggests, the difference might be only a
technical one. The \fhm\ allows us to calculate $n$-current correlators
like the vacuum polarization. Let us take the continuous time limit and
analytically continue to Minkowski space. The full
vacuum polarization can now be reconstructed from its discontinuities.
The discontinuity is a phase space integral of the
product of matrix elements of the source current between
physical states of the \fhm. These matrix elements are still
related to the inverse propagator via the Ward identity.
The pole in the inverse propagator may therefore still reapper in
the vacuum polarization via the matrix elements of the source current,
leading to an inconsistency as before.

  In conclusion, the No-Go arguments as presented in this section are
incomplete. Nevertheless, the failure of a remarkably
diverse range of proposals for \clgt{s} can be understood in the present
framework. Apart from the Smit-Swift model, I refer to the
Eichten-Preskill~\cite{EP}, the waveguide~\cite{wg} and
the staggered fermion~\cite{stgr} models. In all cases, the analytic
structure complies with the above general considerations, and
there is no evidence for zeros in the propagator. This is particularly
clear when weak or strong coupling expansions are available.
See ref.~\cite{noep} for the Eichten-Preskill model, and ref.~\cite{wg}
for the waveguide model.

  In closing this chapter let me mention some recent activity in what is
historically the oldest approach to \clgt{s}. Namely, where the action is
sufficiently non-local to prevent the existence of continuous first
derivatives.

  Slavnov~\cite{slavnov} recently considered the effect of additional
Pauli-Villars fields in a lattice action based on SLAC fermions. He argues
that, at least in perturbation theory, a suitable choice of Pauli-Villars
fields allows one to eliminate the diseases pointed out by Karsten and
Smit~\cite{KS2}.  Another proposal based on SLAC fermions whose details have
not been worked out yet can be found in ref.~\cite{zenkin}.
Finally, an attempt to avoid the doubling precisely
by poles in the inverse propagator was recently made by Bietenholz and
Wiese~\cite{BV}. Their method of ``integrating out of the continuum''
does not have a straightforward diagrammatic interpretation. It is still
an open question whether or not this method leads to a consistent expression
for the vacuum polarization.

\section{DOMAIN WALL FERMIONS}

  I now turn to Kaplan's domain wall fermions~\cite{kaplan}.
I will begin by describing the basic idea in the context
of a vector-like model~\cite{dwdbl}. The reason is that the vector-like
case poses no conceptual problems. Moreover, the use of Kaplan fermions
for lattice QCD minimizes the breaking of axial symmetries by
lattice artifacts~\cite{bndr,mcih,fs,nnv} (see also ref.~\cite{N&N}).
Two proposals for \clgt{s} based on domain wall fermions, the waveguide
model~\cite{wg} and the overlap formula of Narayanan and
Neuberger~\cite{N&N}, will be described next. In this section
I quote only the central results and give very few technical details.
I tried to compensate by giving a rather extensive list of
references. For a review on domain wall fermions see ref.~\cite{dwf}.

  The basic problem with Wilson fermions is the hard breaking of axial
symmetries. The Wilson term has vanishingly small matrix
elements between low energy  {\it free} fermion
states. But the matrix elements of the QCD hamiltonian between
left-handed and right-handed quark states are $O(1)$. As a result, fine tuning
is required in order to obtain the correct continuum limit,
both at the level of the action and in the definition of properly
renormalized operators.

  Kaplan introduces a five dimensional fermion action that contains a Wilson
term, and a mass term which depends explicitly on the fifth coordinate $s$.
The mass function $M(s)$ describes four dimensional {\it defects}, and
these defects support massless four dimensional
chiral fermions. Kaplan originally
used domain wall defects. In a more economic setup the massless chiral
fermions emerge as surface states on the boundaries
of a five dimensional slab~\cite{bndr,mcih}.

  In order to minimize inessential technical details I will stick
to domain wall defects below. In a QCD setup, the lagrangian is
\bqry
  S & = & \sum \{ \bar\j \Sl{D}(U) \j + \bar\j \bo(U) \j \NON
        & &\: + \:  \bar\j \g_5 \partial_5 \j + \bar\j \bo_5 \j
                    + M(s) \bar\j \j \} \,.
\label{dw}
\eqry
The sum runs over the four dimensional coordinates $x_\m$ and the fifth
coordinate $s$, where $-2L < s \le 2L$. Antiperiodic boundary conditions are
assumed in the $s$ direction. The terms involving $\Sl{D}(U)$ and $\bo(U)$ are,
respectively, covariant four-dimensional kinetic and Wilson terms.
The next two terms are (free) kinetic and Wilson terms for the fifth
direction. The last term is the $s$-dependent mass term.
The gauge field that enters \seeq{dw} is {\it four dimensional}. Namely,
$U_{x,s,\m}=U_{x,\m}$ independently of $s$, and $U_{x,s,5}=I$.
The mass function is given by
\beq
  M(s) = \left\{ \begin{array}{ll}
         +M\,, & 0 < s \le 2L\,, \\
         -M\,, & -2L < s \le 0 \,.
         \end{array} \right.
\label{ms}
\eeq
The parameter $M$ (which has nothing to do with the mass of the four
dimensional fermions) is chosen in the range $0 < M < 1$ \cite{N&N}.

  The $s$-dependent mass term defines two defects: a domain wall between
$s=0$ and $s=1$, and an antidomain wall between $s=2L$ and $s=-2L+1$.
In the free field limit, it is easy to check that the five dimensional Dirac
equation has a right-handed and a left-handed zero mode, which are localized on
the domain wall and the antidomain wall respectively. Since the
two zero modes couple to the same gauge field, they actually describe
a single massless quark. A generalization which allows for a non-zero
current mass is discussed in ref.~\cite{bndr,fs}.

  The great advantage of this formulation, is that the
left-handed and right-handed parts of the quark field have
practically disjoint supports in the fifth dimension.
The axially non-conserving part of the
interaction hamiltonian has vanishingly small
matrix elements between any pair of quark states.
As a result, one can rigorously prove~\cite{fs} the restoration of all
non-singlet axial symmetries in the limit $L\to \infty$. In contrast,
the arguments for chiral symmetry restoration are only perturbative
for ordinary Wilson fermions~\cite{KS,it}. At the same time, vectorial
flavour symmetries are manifestly preserved (unlike with staggered fermions).
The proof holds for all values
of the bare coupling $g_0$ and, hence, also in the continuum limit.
It implies that the current mass is only multiplicatively renormalized,
and that operator mixings are restricted by axial quantum numbers as in the
continuum. (In the strong coupling limit the massless spectrum changes.
It is likely that the axial currents defined in ref.~\cite{fs}
become vectorial with respect to the new massless spectrum.)

  The price paid is the necessity of introducing an extra, unphysical
dimension for the fermions.
It is not clear yet how large the extra dimension has to be
in practice, and this question certainly deserves further study.
In order to minimize $L$ one should use the boundary
fermion scheme~\cite{bndr}. If the needed value of $L$ is
sufficiently small, the method may turn out be an attractive alternative for
numerical simulations.

  I now turn to proposals for \clgt{s} based on domain wall fermions. In order
to construct a \clgt\ we should somehow decouple the extra chiral mode at the
antidomain wall. This can be done by introducing the four dimensional
gauge field in the vicinity of the domain wall only. We thus substitute
\beq
   U_{x,s,\m} = \left\{ \begin{array}{ll}
        U_{x,\m}  & -L < s \le L\,, \\
         I\,, & {\rm otherwise},
         \end{array} \right.
\label{wg}
\eeq
in the action \seeq{dw}. The region $-L < s \le L$ is
the ``waveguide'' \cite{wg}. In the limit of large $L$, we clearly succeed in
decoupling the opposite chirality zero mode on the antidomain wall.
However, in going from the previous QCD setup to the waveguide model,
we have created two new defects: these are the interfaces at $s=\pm L$,
where charged degrees of freedom inside the waveguide couple directly to
neutral ones outside of it. As a result,
the fermion action \seeq{dw} with the gauge field \seeq{wg} is not
gauge invariant. The breaking is very mild from the point of view of the
massless modes at the walls, because they are localized far away from both
interfaces. The perturbative philosophy is that mild breaking of gauge
invariance at the level of the lattice action is welcomed, because the
effective action obtained by integrating out the fermions should violate
gauge invariance in the case of an anomalous fermion spectrum. Indeed,
using lattice perturbation theory we recover the consistent
anomaly in the limit of a smooth external gauge field~\cite{dwpert,rs}.

  As we already know, the perturbative analysis may be misleading.
Following the steps of Sect.~2, we can reformulate the waveguide
model in a gauge invariant way, by introducing a Higgs field $V_x$ on the
$s$-links that make the two interfaces. We now clearly see the danger:
If the $V_x$ field is strongly fluctuating (which is true in a symmetric phase)
new massless species (both charged and neutral)
may appear at the waveguide boundaries.
In fact, this is precisely what happens~\cite{wg}. Let us introduce a Yukawa
coupling $y$ that controls the interaction between the $V_x$ field and
the neutral and charged fermions at the interface.
Like the Smit-Swift and Eichten-Preskill models, the phase diagram
of the waveguide model contains two symmetric phases, one at small $y$ and one
at large $y$. The two symmetric phases have different massless spectra.
But in both of them, the new species that appear at the interface render
the spectrum vector-like. In conclusion, the waveguide model fails to yield
a \clgt. Nevertheless, the waveguide model can help us
in studying the physical spectrum of the overlap model, to which I now turn.

  The overlap formula is an ansatz for the fermionic partition function which
was proposed by Narayanan and Neuberger~\cite{N&N} for the construction of
\clgt{s}. This formula can be motivated as follows.
If we consider domain wall fermions on a lattice with an infinite
$s$ direction, there is no antidomain wall and, hence, no unwanted
zero modes with the wrong chirality. Needless to say, one has to construct an
explicit realization of the ``infinite $s$'' situation. The question is
whether the overlap realization evades being the $L\to\infty$ limit
of the waveguide model, or a variant of it. I will return to this question
below.

  The basic observation of ref.~\cite{N&N} is that a
{\it transfer matrix} formalism~\cite{luscher} is particularly powerful
for domain wall fermions.
The reason is that the gauge field is $s$-independent, and so the transfer
matrices that describe the hopping from one four dimensional layer to the
next in the $s$ direction are {\it almost} $s$-independent.
In fact, there are only two different transfer matrices
$T_\pm(U)$ that correspond to the two half-spaces of positive and negative $s$.
The difference between $T_+(U)$ and $T_-(U)$ arises because of the
changing sign of $M(s)$. Formal application of the rules of ref.~\cite{luscher}
suggests that the fermionic partition function, when expressed in a
transfer matrix language, could look something like
\beq
  Z_F(U) \sim \cdots T_-(U) T_-(U) T_+(U) T_+(U) \cdots
\label{inft}
\eeq
The dots indicate that, formally, the products of $T_+(U)$-s and $T_-(U)$-s
continue {\it ad infinitum}.

  In order to arrive at a well defined expression, one observes that in the
limit of large $s$, $T^s_\pm(U)$ projects out the ground state of the
many body hamiltonian $H_\pm(U) = -\log T_\pm(U)$. Explicitly
\beq
  T^s_\pm(U) \to \sket{U\pm} \l_\pm^s \sbra{U\pm} + \cdots
\eeq
Here $\l_\pm$ stands for the largest eigenvalue of $T_\pm(U)$, and
$\sket{U\pm}$ is the corresponding ground state. Focusing on the
interface between positive and negative $s$, one arrives at the following
tentative expression
\beq
  Z_F(U) \sim \braket{U-}{U+} \,.
\label{ovlpfrst}
\eeq
This expression is still not well defined, because states in a Hilbert space
are defined only up to a phase. Adopting the Wigner-Brillouin phase choice,
Narayanan and Neuberger arrive at the {\it overlap formula}
\beq
  Z_{ov}(U) =  { \braket{I-}{U-} \braket{U-}{U+} \braket{U+}{I+}
  \over
  \left| \braket{I-}{U-} \right|\, \braket{I-}{I+}\,
  \left| \braket{U+}{I+} \right|
  }
\label{ovlp}
\eeq
The subscript $ov$ is a shorthand for {\it overlap}.The $\sket{I\pm}$
are the ground states of the free hamiltonians $H^0_\pm=H_\pm(I)$.
Notice that the overlap formula picks only the phase of $\braket{I\pm}{U\pm}$.
The free overlap $\braket{I-}{I+}$ in the denominator is a normalization
factor. \seEq{ovlp} is modified if $\braket{I+}{U+}=0$.
(One can show that  $\braket{I-}{U-}$ never vanishes.) See the last paper
of ref.~\cite{N&N} for the explicit expression.

  In the presence of smooth external gauge fields, the overlap formula
reproduces all the essential properties of chiral fermions.
In particular, the consistent anomaly is recovered in lattice perturbation
theory~\cite{dwpert,rs,N&N}.
More remarkably, the overlap formula vanishes identically in
an instanton background due to level crossing in the spectrum of $H_+(U)$,
and when the correct number of fermionic creation
(annihilation) operators is inserted, instanton induced transition amplitudes
are reproduced as well~\cite{N&N}. Finally, by multiplying the overlap
formulae for an equal number of left-handed and right-handed fermions,
one arrives at a valid model for lattice QCD~\cite{nnv}.

  All implementations of Kaplan's domain wall fermions require
{\it subtractions} to cancel undesirable effects of the infinitely many
four dimensional fields with cutoff mass. These subtractions can be
represented as five dimensional Pauli-Villars (PV) fields~\cite{frsl}.
In perturbation theory, the only difference between the overlap formula
and the waveguide model is in the choice of the PV fields.
The PV contribution to the effective action (the sum of the one loop diagrams)
is purely {\it real}. But the perturbative ``signature'' of a chiral fermion
is the {\it imaginary} part of the effective action. The overlap formula and
the waveguide model have equal imaginary parts for their effective actions
to all orders in lattice perturbation theory~\cite{dwpert}.
This is the first sign that the
overlap and waveguide models may be closely related, and that the overlap
model may be vulnerable to the same dangers that ultimately
render the waveguide model vector-like. It is true that the two models
differ in the presence of gauge fields whose total topological charge is
non-zero~\cite{nnnew}. However, as I will now explain, the significance
of this observation is limited.

  Following the reasoning of sect.~2, the spectrum of the overlap model
is determined by the \fhm\  defined by the partition function
\beq
  Z'_{ov} = \int \cd V\, Z_{ov}\Big(U=I^{(V)}\Big)
\label{ovlph}
\eeq
The pure gauge configurations, whose nonperturbative dynamics determines
the spectrum of the overlap model, have {\it zero} total topological charge.
As in Sect.~2, one should first find whether the global symmetry of $Z'_{ov}$
is spontaneously broken. If it is not, one should proceed to calculate the
fermion spectrum. To this end, one should augment \seeq{ovlp} by giving its
dependence on an appropriate set of external sources.

  The partition function $Z'_{ov}$ does not arise from a local
action. This makes its investigation particularly difficult. When specifying
the operators that couple to the external sources one has to use some
physical intuition, and it is not easy to tell whether extra massless state
are hiding somewhere. If no new massless states are generated and the
spectrum is truly chiral, one would expect {\it zeros in the propagator}.
One would then have to check whether or not these zeros imply some
inconsistency. A less dramatic possibility~\cite{ysnogo} is that the
spectrum is actually vector-like, and the zeros in the propagator
simply reflect an {\it under complete} set of external sources
(or $\c$-s in the language of Sect.~2). This could happen if
the chosen external sources do not create the new, dynamically generated,
massless fermion. The question could be settled in principle by calculating
$n$-current correlators within the \fhm\ (the vacuum polarization and more) and
inferring the spectrum from them. But, in practice, this
strategy may be hampered by technical complications.

  A way to circumvent the above difficulty is to show that, at least
in some special cases, the partition function $Z'_{ov}$ does have
an interpretation as a local field theory.
The main result of ref.~\cite{gs1} (which corrects an error in a previous
publication~\cite{gs2}) is the following.
If the target \clgt\ contains $4n$ chiral families, then for {\it all} $V_x$
\beq
  Z_{ov} \Big( I^{(V)} \Big) = Z_{m.w.g.} \Big( I^{(V)} \Big)\,.
\label{equal}
\eeq
$Z_{m.w.g.}(U)$ is the fermionic partition function of a {\it modified
waveguide model} with a Yukawa coupling $y=1$.
\seEq{equal} implies that the overlap model and the
modified waveguide model lead to {\it the same} \fhm. This, in turn,
implies that the two models have the same spectrum.

  The phase diagram and spectrum of the modified waveguide model can be
studied using previously developed techniques~\cite{wg,gs2}.
The difference between the original and modified waveguide
models is in the choice of the PV fields. At least for small $y$,
one expects that both variants of the waveguide model
(and, hence, the overlap model too) will have the same {\it vector-like}
fermion spectrum. (The modified waveguide and overlap models will also
have massless ghost states.) Another interesting question is
whether the equivalence between the overlap formula and some waveguide model
can be extended to topologically nontrivial sectors as well.
See ref.~\cite{gs1,gs2} for more details.

  Domain wall fermions were also investigated using a five dimensional (5-d)
gauge field~\cite{dwfiv}. This approach does not produce a \clgt. Depending on
the gauge coupling in the $s$ direction, the 5-d gauge field is reduced
dynamically to a 4-d gauge field (see the first paper of ref.~\cite{wg}), or
else the system breaks up into an infinite collection of independent 4-d
theories of Wilson fermions, the so-called layered phase.

\section{GAUGE FIXING}

  The No-Go arguments of Sect.~2 can be circumvented by imposing
{\it constraints} on the lattice gauge field's {\it measure}. This procedure
is usually called ``gauge fixing'', although this term
is highly inaccurate in the present context. In lattice QCD
the action is gauge invariant, and gauge fixing means picking a
single representative out of many equivalent ones.
But in trying to construct \clgt{s} one uses {\it gauge variant}
fermion actions. In this case, imposing constraints on the gauge field's
measure is really a part of the definition of the theory.

  The gauge fixing approach was originally proposed by the Roma
group~\cite{roma}, and subsequently followed by the Zaragoza group~\cite{zrgz}.
So far, the results are limited to the context of lattice perturbation
theory. In this section I will present considerations
which are relevant for a nonperturbative implementation of this
approach. The main conclusion is that one needs a {\it global} gauge fixing
algorithm that selects only
relatively {\it smooth} lattice gauge field configurations. This diagnosis
has already been made by Vink~\cite{vink}. Vink proposed a method that
effectively couples the fermions only to maximally smooth gauge field
configurations. His method, however, has not been investigated in any
detail so far.

  Consider the perturbative effective action
$S_{\rm eff}=S_{\rm eff}(U)$ defined as the
sum of the one loop diagrams on the lattice. If the fermion action is not
gauge invariant, the first variation of $S_{\rm eff}$ with respect to a
lattice gauge transformation parametrized by $g_x=\exp(i\o_x)$ will look like
\beq
   {\d S_{\rm eff} \over \d \o_x} =
   c_0\, \ca_x + \sum_{n\ge 1} c_n a^n \co_x^{(n)}
\label{an}
\eeq
$\ca_x$ is some discretized version of the consistent anomaly, and the
coefficient $c_0$ will vanish if the perturbative spectrum is anomaly free.
In writing \seeq{an} I assume that all other operators of
dimension less than or equal to four have been cancelled by counter terms.

  As the infinite sum on the \rhs of \seeq{an} indicates,
$\d S_{\rm eff} / \d \o_x$ contains more than just
the consistent anomaly. In fact, $\ca_x$ is the first term in an
infinite series in the lattice spacing $a$. The $\co_x^{(n)}$ are local
operators of dimension $n+4$. Like $c_0$, the coefficients $c_n$, $n>0$ have
a group theoretical origin. All the coefficients will
vanish simultaneously {\it iff} the fermion action is exactly gauge invariant.
Unless the fermion action is highly non-local, this implies that already the
perturbative spectrum is vector-like.

  If the perturbative spectrum is chiral but {\it anomaly free},
$\d S_{\rm eff} / \d \o_x$ will vanish in the limit of
smooth external gauge fields. To see this, we observe that all the
$\co_x^{(n)}$ must contain lattice derivatives. Non-derivative terms can
be calculated in the limit of zero external momentum, and so they should
agree with some continuum regularization. Namely, they should give rise to
the non-derivative part of the gauge variation of counter terms,
and nothing more. \seEq{an} is therefore effectively an expansion in $a p_\m$.
(This property becomes manifest in explicit computations based on standard
momentum space Feynman rules.) In the adiabatic limit, the infinite series
in \seeq{an} tend to zero, because the external field contains only
Fourier modes with $a p_\m\to 0$.

  The difficulty arises because the lattice momentum is not conserved
under gauge transformations. If we apply a generic lattice gauge
transformation to a smooth gauge field, we will find that the typical
momentum in the transformed configuration is $ap_\m \sim 1$. Substituting
in \seeq{an} we see that the the gauge variation
of the effective action is now $O(1)$.
Thus, even a formally ``mild'' breaking of gauge invariance by the
fermion action is really very large for a generic lattice gauge field
configuration. This large generic breaking is in conflict with the fact that
the continuum limit must describe a gauge invariant theory.
If a non-trivial continuum limit exists, gauge invariance has to be
restored {\it dynamically}. Sect.~2 explain why the
dynamical restoration of gauge invariance comes at the price of
producing a vector-like spectrum.

  The above considerations suggest that the aim of the ``gauge fixing''
procedure should be to keep only relatively smooth lattice gauge field
configurations, while excluding all the rest.
The restriction to relatively smooth configurations clearly reduces the
breaking of gauge invariance, as represented by the \rhs of \seeq{an}.
If we intend to adopt this strategy, we first need some understanding of the
structure of the lattice gauge orbit. Let me begin by considering
an arbitrary configuration of the lattice gauge field, and some algorithm
that rotates the configuration to one that satisfies the Landau
gauge condition. The question is how will the rotated configuration
look like. To answer that question, we observe that the Landau gauge
condition is satisfied by extrema of the functional
$\cf(U)=\Re \sum_{x,\m} \tr U_{x,\m}$ along the orbit.
The structure of the rotated configuration will therefore depend strongly
on whether we have arrived at a local or a global maximum of $\cf$.

  Recent work reveals a proliferation of solutions to the Landau gauge
condition (Gribov copies) on the lattice~\cite{fix}. The basic reason is
that the lattice is a {\it multiply connected} topological space.
Consider for definiteness a $U(1)$ theory in two dimensions, and let
$\exp(i\th_x)$ be the gauge transformation that takes us from one
Gribov copy to another. If $\th_x$ was an ordinary function, it would have
to satisfy the lattice Laplace equation. The scalar laplacian
has no non-trivial solutions, which is why an abelian theory is free of
Gribov copies in the continuum.

  However, $\th_x$ is a periodic variable. In other words, a lattice
gauge transformation is a mapping from the lattice sites to the unit circle.
If we extend it to a mapping $\Th(\bo)$ from the perimeter of the plaquette
to the unit circle, we will find that $\Th(\bo)$ can be classified
according to its homotopy class. Notice that some choice has to be
made in the definition of $\Th(\bo)$.
When interpolating between $x$ and $x+\hat\m$, we will choose
to go along the unit circle in the direction that minimizes the arc length
between $\exp(i\th_x)$ and $\exp(i\th_{x+\hat\m})$~\cite{wind}.
This procedure allows us to asign a {\it local winding number} to the gauge
transformation. A winding number $n=\pm 1$ means that the gauge
transformation creates a singular (anti)vortex inside the plaquette.
(Periodic boundary conditions will force equality of the total
number of vortices and antivortices.)

  The main result of ref.~\cite{fix} is that lattice Gribov copies are in
one to one correspondence with singular vortices.
Fig.~6 of ref.~\cite{fix} describes a gauge fixed
configuration obtained as follows. In the first step, a random lattice
gauge transformation was applied to a smooth configuration.
In the second step, a standard {\it local} algorithm was applied to
enforce the Landau gauge condition. A glance at Fig.~6
reveals islands of large $A_\m$ with smooth $A_\m$ in between,
supporting the picture that the Gribov copies arise due to a
localized structure. The conclusion is that local extrema of the Landau
gauge functional $\cf$ are characterized by a vortex-antivortex gas
with {\it finite density} in lattice units.

  Large values for $\co_x^{(n)}$ in \seeq{an} are correlated
with large gradients in $U_{x,\m}$. Such large gradients will exist in the
vicinity of every singular vortex. Consequently, in order to minimize
the \rhs of \seeq{an} we have to eliminate all the Gribov copies.
This, in turn, requires the use of a {\it global} method.
(See ref.~\cite{vink,fix} for specific examples.)
It is an open question whether
one can construct a global algorithm that reduces the breaking of gauge
invariance  sufficiently, without at the same time spoiling other
important properties of the continuum limit.

\section{INTERPOLATING FIELDS}

  An alternative approach to the definition of chiral gauge theories is
to put the gauge field on the lattice, while keeping the fermions in the
continuum. Interest in this approach arose following a recent paper by
't Hooft~\cite{thooft}. (See ref.~\cite{js,interl,topo} for earlier relevant
works.) A central element of the method is the construction of a continuum
interpolation of the lattice gauge field~\cite{intrp}.
The second element is a nonperturbative
definition of the chiral fermion determinant for any continuum gauge field
that can be obtained via the interpolation. The interpolating fields method may
be problematic to implement in numerical simulations. Here I will only
discuss the conceptual question of whether the method can provide
a consistent definition of chiral gauge theories.

  For definiteness, I will assume that the continuum chiral Dirac operator
is $\hat{D} = \sl\partial + i\Sl{A} P_L$. The basic property of
$\hat{D}$ is that its eigenvalues $\l_i$ are complex and {\it gauge variant},
and that a right eigenstate defined by $\hat{D}\j_i=\l_i\j_i$ is not the
complex conjugate of the corresponding left eigenstate $\c_i\hat{D}=\c_i\l_i$.

  The obvious reason why one may hope that the interpolating fields method
will do better than pure lattice approaches, is that there is no fermion
doubling in the continuum. As a first step, let us see in what way
this new method may change the considerations
of the previous section. While the gauge field is still regularized by
the lattice cutoff, we need a separate
regularization to define the continuum fermion determinant.
This regularization introduces a new cutoff scale $M$, which now controls the
violations of gauge invariance. Assuming $aM\gg1$ and using the same operator
basis as before, one expects that \seeq{an} will be replaced by
\beq
   {\d S_{\rm eff} \over \d \o_x} =
   c_0\, \ca_x + \sum_{n\ge 1} c'_n M^{-n}  \co_x^{(n)}
\label{ann}
\eeq
The coefficient $c_0$ of the (discretized) consistent anomaly is of course
the same as in \seeq{an}.
The main change is that \seeq{ann} represents an expansion in $p_\m/M$.
Remember that the generic momentum of a lattice gauge field is $O(1/a)$.
The new expansion parameter is therefore effectively $1/(Ma)$.
If we now send $M\to\infty$ at a fixed $a$,
we may hope that the \rhs of \seeq{ann} will tend to zero
for an anomaly free spectrum. (To avoid confusion, let me stress
that the above heuristic considerations are based on
perturbative reasoning and, in any event, they are no substitute for
a detailed proof.)

  Let me begin with a list of features that we expect from the interpolating
field. There are three fundamental requirements.
First, the interpolating field $A_\m(x)=A_\m(x;U)$ should reproduce every
lattice link variable $U_{\vec{n},\m}$ by calculating the parallel transporter
along that link from $\vec{n}$ to $\vec{n}+\hat\m$. (In this section I use
$\vec{n}$ to denote a lattice site.) The second property is
{\it gauge covariance}. Consider two lattice gauge fields which
are related by a lattice gauge transformation
$U'_{\vec{n},\m}= g_{\vec{n}} U_{\vec{n},\m} g^\dagger_{\vec{n}+\hat\m}$.
We demand that the corresponding interpolating fields will be related
by a continuum gauge transformation
\beq
  A'_\m(x)=\O(x)(A_\m(x)-i\partial_\m)\O^\dagger(x) \,,
\label{aap}
\eeq
where $A'_\m(x)=A_\m(x;U')$, and $\O(x)=\O(x;g,U)$ is a
continuum interpolation of the lattice gauge
transformation, that coincides with $g_{\vec{n}}$ at the lattice points.
(Notice that $\O(x)$ can depend on the $U_{\vec{n},\m}$-s too).
The third requirement, which stems from the need to have a well behaved
spectrum for $\hat{D}$, is that the worst singularities in $F_{\m\n}(x;U)$
are discontinuities. (This requirement is weaker than
``transversal continuity'' which is often mentioned in the literature.)

   I now turn to specific interpolation techniques.
A method  based on a linear interpolation kernel is discussed in
ref.~\cite{linint}. The method is very simple, but it is
highly specific to {\it non-compact} U(1). Here I will focus on the
interpolation proposed by G\"ockeler {\it et al}~\cite{intrp}.
I will describe the method for {\it compact} U(1) in two
dimensions. The reason is that the formulae are much simplified in this
case, which still contains all the essential (and in particular topological)
features of non-abelian theories in four dimensions.

  In two dimensions, we have to define the continuum gauge field first
on the links, and then inside each plaquette. We define the interpolating field
to be constant on all points that make a given link $\{\vec{n},\m\}$.
Explicitly, for $\vec{x}=\vec{n}+t\hat\m$, $0\le t < 1$, one defines
\beq
  A_\m(\vec{x})=A_\m(\vec{n})=i\log U_{\vec{n},\m}\,.
\label{amu}
\eeq
The logarithm is always taken such that $|A_\m(\vec{n})|<\p$.
The interpolating field is left undefined on
a zero measure subset of lattice gauge fields,
where $U_{\vec{n},\m}=-1$ for some links.

  We next have to extend the interpolation inside a given plaquette,
which we parametrize as $\vec{x} = \vec{n} + \vec{t}$, where
$0\le t_1,t_2 \le 1$. As a first step, let us assume that the link variables
satisfy a local axial gauge, where $U'_{\vec{n}+\hat{1},2}$ is the only
non-trivial link variable. In this special case,
the interpolating field is given by
\bqry
  A'_1 & = & 0 \,, \NON
  A'_2 & = & it_1 \log U'_{\vec{n}+\hat{1},2} \,.
\label{ax}
\eqry
This leads to a constant field strength throughout the plaquette,
given by $F = i\log U'_{\vec{n}+\hat{1},2}$. The generalization
to arbitrary values of the link variables is done as follows.
The {\it lattice gauge transformation} that enforces the local axial
gauge is first extended to a {\it continuum gauge transformation}
$\O(\vec{x}\,)$ throughout the plaquette. The interpolating field
$A_\m(\vec{x}\,)$ is then defined by inverting \seeq{aap}.

  Let us compare the field strength $F$ to the directed sum along the
perimeter of the $A_\m$-s defined in \seeq{amu}. In general, we may find
\bqry
  F & = & A_1(\vec{n}) + A_2(\vec{n}+\hat{1}) \NON
    & & - A_1(\vec{n}+\hat{2}) - A_2(\vec{n}) + 2\p k \,.
\label{F}
\eqry
Here the possible values of $k$ are $k=-2 \ldots 2$, and
the value changes whenever a link variable goes through $-1$.
Consider now the restriction $\O|_\Box=\O|_\Box(\vec{z}\,)$ of the continuum
gauge transformation to the perimeter of the plaquette. In fact,
$\O|_\Box(\vec{z}\,)$ is completely determined by
$A_\m(\vec{n})$ and $A'_\m(\vec{n})$ via the
{\it gauge covariance condition} \seeq{aap}
(up to a constant overall phase). On each link, the phase
of $\O|_\Box(\vec{z}\,)$ varies linearly. For the three links where
$A'_\m(\vec{n})=0$, this construction coincides with the minimal arc
prescription of the previous section. But in general this is not true for
the link where  $A'_\m(\vec{n})\ne 0$.
As a mapping into the unit circle, $\O|_\Box(\vec{z}\,)$ is characterized
by a {\it winding number}. Comparing \seeq{F} to the line integral of
\seeq{aap} along the perimeter, we conclude that
the winding number of $\O|_\Box(\vec{z}\,)$ is $k$. (A winding number will
arise in four dimensional non-abelian theories too, when one extends the
gauge transformation to the {\it faces of the hypercube}.)

  The extension of $\O(\vec{x}\,)$ from the perimeter to the entire plaquette
depends crucially on the winding number. For $k=0$, one can extend the phase
of $\O(\vec{x}\,)$ linearly, resulting in the interpolating field
\bqry
  A_1 & = & (1-t_2) A_1(\vec{n}) +
                     t_2 A_1(\vec{n}+\hat{2}) \,, \NON
  A_2 & = & (1-t_1) A_2(\vec{n}) +
                     t_1 A_2(\vec{n}+\hat{1}) \,.
\label{k0}
\eqry
Notice that \seeq{k0} is rotationally covariant.

  If the local winding number $k=k(\vec{n})$ vanishes everywhere, the
resulting interpolating field is also transversally
continuous. Namely,  $A_\m(x;U)$ is continuous going across a plaquette
boundary in the $\n$ direction for $\n\ne \m$, but in general not for
$\m=\n$. The longitudinal discontinuities lead to an enhanced
content of high momentum modes in $\tilde{A}_\m(q;U)$.
This may have undesirable effects on the UV behaviour~\cite{uv}.
However, the UV behaviour can be improved by local smoothing of the
interpolating field~\cite{intrp}.

  If $k(\vec{n})\ne 0$, there has to be an (anti)vortex singularity inside
the  plaquette $\vec{n}$. (Similarly, in four dimensions $k(\vec{n})\ne 0$
implies an (anti)instanton singularity inside the hypercube.)
The formulae of ref.~\cite{intrp} lead to
a {\it line discontinuity} in $\O(\vec{x}\,)$.
If, say, the line is horizontal, the resulting $A_2(\vec{x}\,)$ will have
a $\d$-function piece localized on that line, while $A_1(\vec{x}\,)$ will
not be transversally continuous across it. The two ugly looking
singularities cancel of course in $F$.

  A different extension of the gauge transformation can be defined by
demanding that  $\O(\vec{x}\,)$ be constant along each ray emanating from
the center point $\vec{x}_0$.  Explicitly,
let $\vec{x}=t\vec{x}_0 + (1-t)\vec{z}$, where $\vec{z}\in \bo$ is a boundary
point and $0 \le t < 1$, then we define $\O(\vec{x}\,)=\O|_\Box(\vec{z}\,)$.
The resulting singularity in $A_\m(x)$ is now very similar to that of a
continuum vortex.

  In summary, the interpolation leads to a well defined quasi-local topological
structure, which is characterized by the {\it disorder field} $k(\vec{n})$.
For $k(\vec{n})\ne 0$, both the method of ref.~\cite{intrp} and the one
described above are not rotationally covariant. This is not a serious
problem, however,  because rotational covariance can always be restored by
averaging over different choices for the direction of the local axial gauge.

  The real problem is how to define the continuum fermion determinant
nonperturbatively such that, in an anomaly free theory, it will be gauge
invariant (up to local counter terms) for {\it all} interpolating fields.
Using Pauli-Villars regularization, 't~Hooft~\cite{thooft} proves gauge
invariance in the limit $Ma\to\infty$
under the assumption that $A_\m(x;U)$ is {\it globally bounded}.
(For other proofs see ref.~\cite{prv}.)
But $A_\m(x;U)$ is globally bounded {\it iff} the corresponding
disorder field $k(\vec{n})$ vanishes everywhere. Notice that
interpolating fields with $k(\vec{n})=0$ are highly non-representative.
In a generic interpolating field, a {\it finite
fraction} of all plaquettes (hypercubes in four dimensions) will carry a
non-zero winding number. Thus, in the thermodynamical limit, the proof of
ref.~\cite{thooft} covers only a vanishingly small subset of the
interpolating field's space.

   In a similar spirit to the previous section, non-zero values for
$k(\vec{n})$ can be suppressed (or even eliminated
altogether for non-periodic boundary conditions) if one adopts a global
method for constructing the interpolating fields. An inspection of \seeq{F}
reveals that enforcing $k(\vec{n})=0$ has a price. Namely, unlike in
\seeq{amu}, one has to allow the interpolating field on the links
$A_\m(\vec{n})$ to lie outside of the interval $(-\p,\p)$.
In the infinite volume
limit, we still cannot establish the existence of a global bound on
$|A_\m(x)|$. The reason now is that the magnitude of the
so constructed interpolating field may be infra-red divergent.

\section{CONCLUSIONS}

  The power and beauty of lattice QCD stems from its manifest gauge invariance.
In trying to understand why it is so difficult to construct \clgt{s} one
naturally focuses on the properties of the lattice {\it fermions}.
In this review I have stressed the complementary
role of the lattice {\it gauge field}.

  The crucial feature is that the transversal degrees of freedom
described by the lattice gauge field always couple to a {\it conserved current
defined on the lattice links}. Only the lattice gauge field's measure
is used in establishing this property. This allows us to apply the No-Go
theorems to an effective hamiltonian, constructed in a natural way from
the inverse two point function of the massless fermions. The most important
open question is whether {\it poles in the inverse two point function}
can evade the No-Go theorems consistently.

  By invoking continuum fermions, the interpolating fields method seems to
evade the reasoning of Sect.~2. The real situation, however,
is more subtle. Loosely speaking, although our fermions are defined in the
continuum, what the lattice gauge field can feel is only an effective fermion
field that lives on a lattice too. In more detail, the ultimate role of the
interpolation is to define  the fermion determinant, or the fermionic
partition function, as a functional of the lattice gauge field.
The variation of the fermionic partition function with respect to
$U_{x,\m}$ defines the source current associated with the same link.
Since the current lives on lattice links, one expects that its
matrix elements could be reproduced by some lattice action. That action
could be extremely complicated to write down, but the only thing
we need to know is that it exists, and that it is very mildly local.
If true, we can apply the reasoning of Sect.~2 to the
interpolating fields method too.

  Logically, a way out is to {\it constrain} the lattice gauge field's
measure. In Sect.~4 I discussed at a heuristic level, how the
restriction to relatively smooth gauge fields reduces
violations of gauge invariance. As of today, very little is known on this
approach. The first question is whether a global algorithm can
suppress violations of gauge invariance {\it sufficiently}.
Ref.~\cite{bhs} finds that
{\it nonperturbative counter terms} might be needed to completely
restore gauge invariance. In my opinion, nonperturbative counter terms are
equivalent to an infinite amount of fine tuning. We should therefore
hope that we can do without them. Also,
the use of any global algorithm introduces some non-locality into the
theory. One should check whether this does not spoil some
important property of the continuum limit, such as causality or
unitarity.

  Another issue that I have not addressed in this review, is the fact that
proposals for chiral lattice gauge theories tend
to have an exactly conserved U(1) symmetry associated with fermion number.
This feature leads to an apparent conflict with fermion number
non-conservation~\cite{bd}. On the other hand, there are arguments
that the paradox can be resolved~\cite{dm,bhs}. The investigation
of all these important questions should clearly continue in the future.

  I thank Jim Hetrick, Karl Jansen and, especially, Maarten Golterman
for many discussions and for very useful comments.




\end{document}